# Hybrid-Integrated DFB-Laser-Coupled 1 × 8 Thin-Film Lithium Niobate Modulator Array for High-Speed Parallel Optical Transmitters

*Qiyue Hu,*[1,2] *Junxia Zhou,*[2,*] *Zhe Wang,*[2] *Botao Fu,*[2] *Jinming Chen,*[2] *Yunpeng Song,*[2] *Mengqi Li,*[2] *Dewei Zhang,*[1,2] *Yuheng Chen,*[1,2] *Jinxin Huang,*[1,2] *Min Wang,*[2] *Jia Qi,*[2,*] *Ya Cheng,*[1,2,3,4,5,*]

[1]State Key Laboratory of Precision Spectroscopy, East China Normal University, Shanghai 200062, China.

[2]The Extreme Optoelectromechanics Laboratory (XXL), School of Physics, East China Normal University, Shanghai 200241, China.

[3]Hefei National Laboratory, Hefei 230088, China.

[4]Shanghai Research Center for Quantum Sciences, Shanghai 201315, China.

[5]Collaborative Innovation Center of Extreme Optics, Shanxi University, Taiyuan 030006, China.

*Corresponding Author: E-mail: jxzhou@phy.ecnu.edu.cn; jqi@phy.ecnu.edu.cn; ya.cheng@siom.ac.cn

Keywords: thin-film lithium niobate, electro-optic modulator array, hybrid laser integration, parallel optical interconnects

**Abstract:** Thin-film lithium niobate (TFLN) electro-optic modulators are attractive for high-speed optical interconnects, but scalable transmitter architectures require not only high modulation bandwidth but also multi-channel optical power distribution and practical laser-to-chip integration. Here, we demonstrate a hybrid-integrated 1 × 8 TFLN electro-optic modulator array passively butt-coupled to a 1550 nm distributed-feedback laser (DFB). The chip integrates a three-stage cascaded 1 × 2 multimode-interference splitter, spot-size converters, eight traveling-wave Mach-Zehnder modulators, thermal tuning electrodes, and on-chip 50 Ω terminations. The cascaded splitter provides uniform optical power distribution with a maximum normalized power deviation of 9.7%, while the optimized electrodes enable electro-optic 3 dB bandwidths exceeding 40 GHz for all channels. The measured half-wave voltages are 3.60-3.83 V, corresponding to half-wave voltage-length product of 2.52-2.68 V·cm for a 7 mm modulation length, and the extinction ratio reaches approximately 25 dB. The bare-chip insertion loss is 15.19-16.55 dB, and DFB laser bonding introduces an additional coupling loss of approximately 5 dB while preserving channel uniformity. These results establish a practical

TFLN-based multi-channel modulator platform and represent a step toward compact hybrid-integrated optical transmitters for high-speed parallel interconnects.

**1. Introduction**

The rapid growth of artificial intelligence, cloud computing, and high-performance computing is increasing both the aggregate capacity and energy-efficiency requirements of data-center optical interconnects. This trend motivates optical transmitter technologies that can support high channel counts, broad analog bandwidth, low insertion loss, and packaging-compatible integration with electronic and laser sources[1-3]. In this context, thin-film lithium niobate (TFLN) has emerged as a compelling integrated-photonics platform because it combines the large Pockels coefficient, broad transparency window, low optical loss, and high power handling of lithium niobate with lithographically defined waveguides and scalable wafer-level processing[4-7].

Single-channel TFLN modulators have reached an advanced level of maturity. Integrated Mach-Zehnder and related device architectures have demonstrated CMOS-compatible drive voltages, low optical loss, large electro-optic bandwidth, and high-speed data transmission[8-19]. These results establish TFLN as one of the most competitive platforms for high-speed optical modulation. For practical optical engines, however, the key challenge is shifting from single-device performance to array-level integration: the optical input must be distributed uniformly to multiple channels, the channel-to-channel modulation characteristics must be well matched, and the light source must be coupled to the modulator chip by a process compatible with manufacturable packaging.

Several issues are particularly important for multi-channel TFLN transmitter arrays. First, the on-chip 1 × N power distribution network must be low-loss, fabrication-tolerant, and uniform; otherwise, the extinction ratio, optical signal-to-noise ratio, and link margin will vary across the array[20]. Second, although direct heterogeneous integration of III-V gain media on lithium niobate is an attractive long-term direction, it remains technically demanding because of material mismatch, thermal-budget constraints, fabrication complexity, and cost. A more immediately practical route is to combine mature C-band distributed-feedback (DFB) lasers with TFLN modulator chips by passive edge coupling, an approach that can preserve the advantages of commercial semiconductor lasers while exploiting the high-speed modulation capability of TFLN[21, 22]. Third, an experimentally verified transmitter chain from laser input to on-chip splitting, parallel modulation, and high-frequency characterization remains necessary for evaluating the practical scalability of TFLN arrays.

Here, we report a hybrid-integrated 1 × 8 high-speed electro-optic modulator array on a TFLN platform. The chip monolithically integrates a three-stage cascaded 1 × 2 multimode-interference (MMI) splitter, eight traveling-wave Mach-Zehnder modulators (MZMs), spot-size converters (SSCs), thermal tuning electrodes, and on-chip 50 Ω terminations. Experimentally, all eight channels exhibit electro-optic 3 dB bandwidths above 40 GHz and half-wave voltages of 3.60-3.83 V, corresponding to $V_{\pi}L$ of 2.52-2.68 V·cm. By passively butt-coupling a 1550 nm DFB laser to the chip, we further demonstrate a hybrid-integrated eight-channel electro-optic light source and quantify the insertion-loss penalty introduced by laser bonding. This work addresses the array-level integration problem of TFLN modulators and provides a practical route toward compact multi-channel optical transmitters for high-speed parallel interconnects.

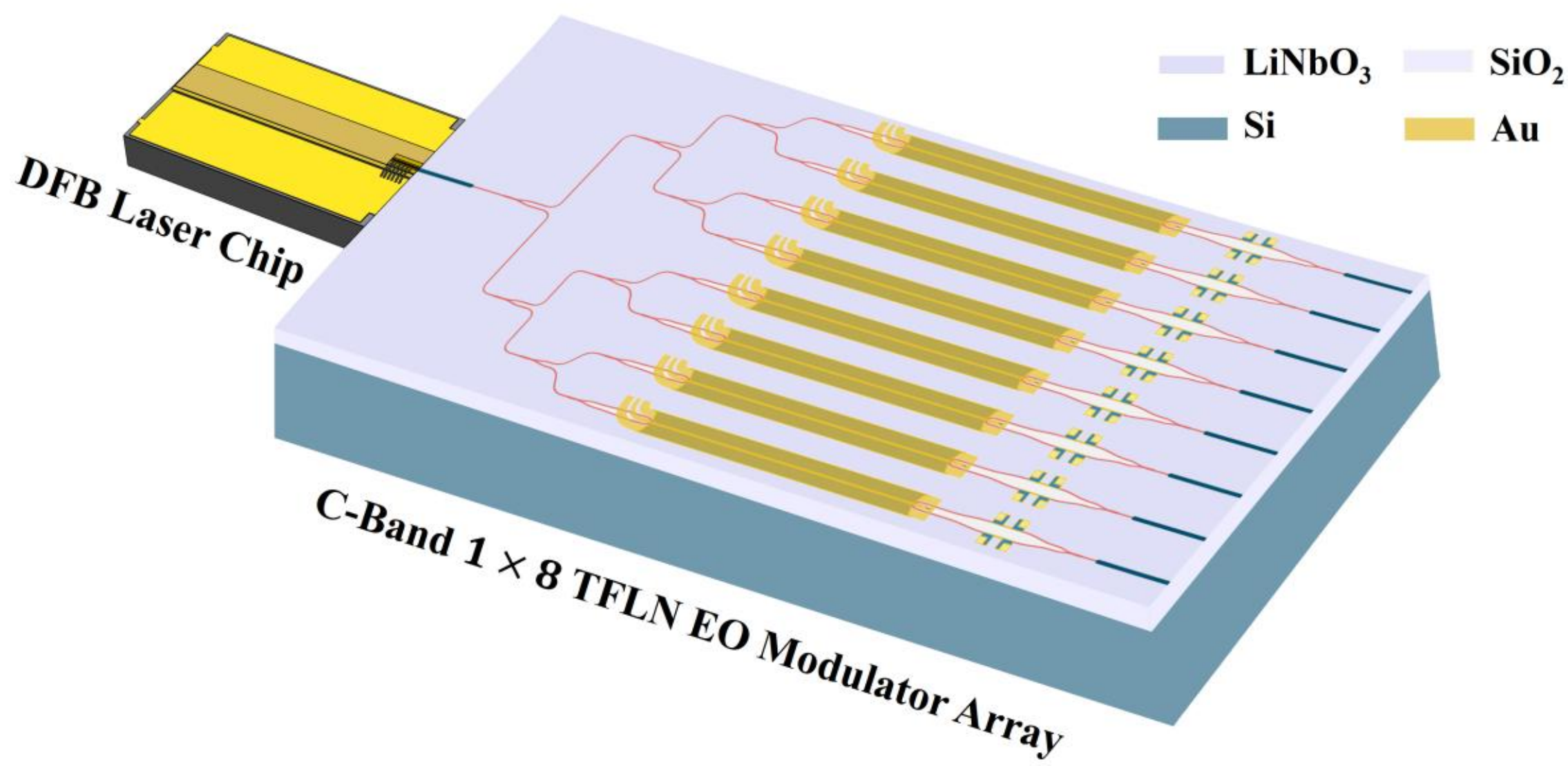

**Figure 1.** Schematic structure of the C-band 1 × 8 TFLN electro-optic modulator array chip. A cascaded 1 × 2 MMI network distributes the input optical power into eight arms, each integrated with a traveling-wave Mach-Zehnder modulator.

## 2. Device Design and Simulation

Figure 1 illustrates the overall architecture of the C-band 1 × 8 TFLN electro-optic modulator array. A cascaded MMI network is used to provide uniform power splitting, and each output arm is connected to an MZM driven by optimized traveling-wave electrodes. Compared with a single-stage 1 × 8 MMI, the cascaded 1 × 2 topology reduces sensitivity to dimensional errors and wavelength drift. It also provides a modular layout in which the splitter and modulator sections can be independently optimized. This architecture is well suited for multi-channel optical transmitters because it combines compactness, parallelism, and fabrication tolerance.

### 2.1. 1 × 8 Uniform Optical Power-Splitting Network

A low-loss and channel-uniform splitter is essential for consistent parallel modulation. In this design, a three-stage cascaded 1 × 2 MMI network is adopted to form the 1 × 8 optical power splitter. The width and length of the multimode region, the coupling-section position, and the input/output waveguide offsets are optimized through parameter sweeps. Bézier-curve bends are used in the routing waveguides to suppress high-order-mode excitation and radiation loss. The optimized single-stage MMI has a multimode region with a width of 16 µm and a length of 195.4 µm. At 1550 nm, the simulated insertion loss is 0.026 dB and the power-splitting ratio is close to 1:1, as shown in Figure 2a. The tolerance analysis in Figure 2b indicates that the transmittance remains above 0.49, corresponding to an insertion loss below 0.087 dB, under width variations of ±100 nm and length variations of ±3 µm. The simulated cascaded 1 × 8 network maintains stable transmission over 1500-1600 nm, with a theoretical insertion loss of approximately 0.084 dB at 1550 nm (Figure 2c).

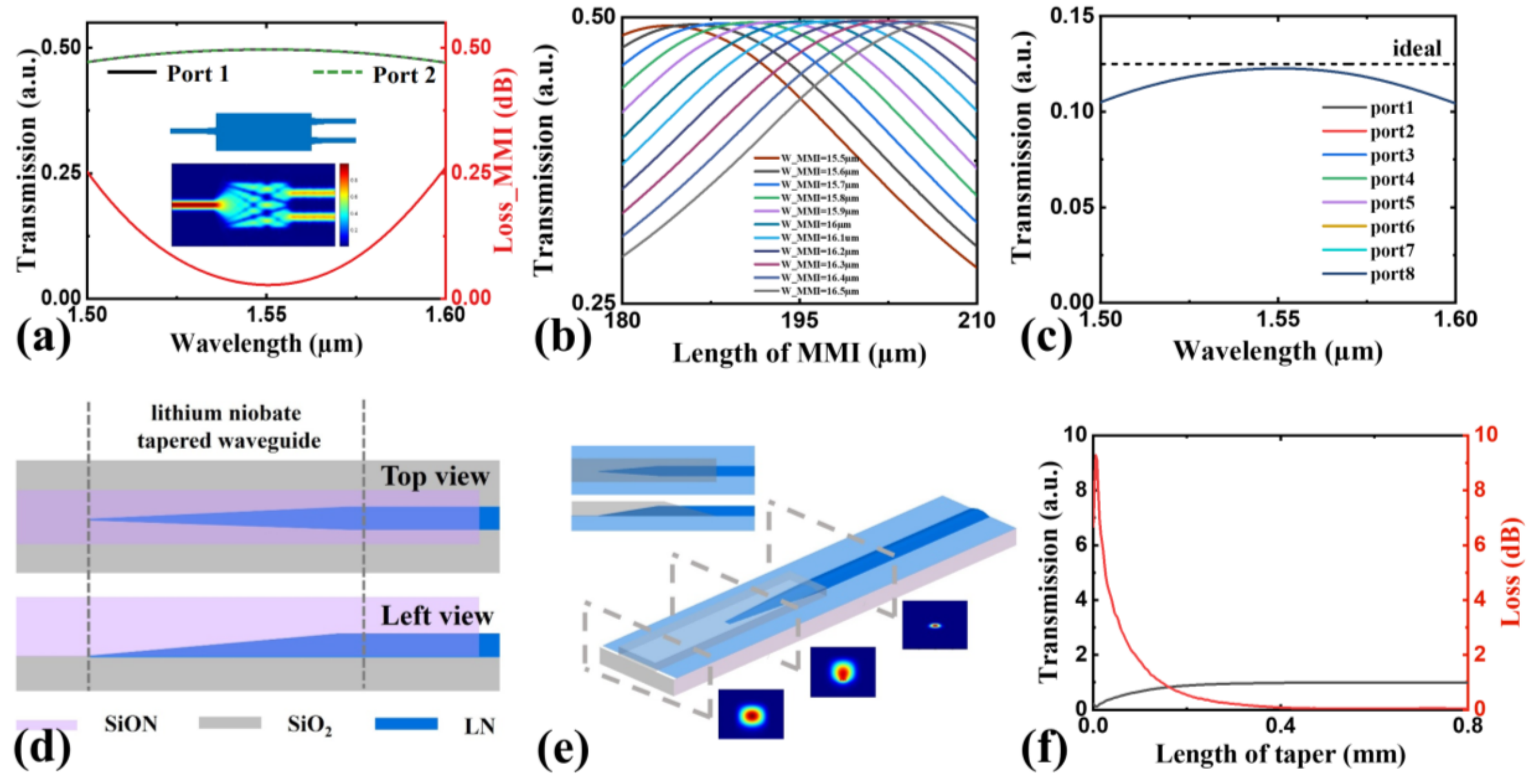

**Figure 2.** Design and simulation of the optical power splitter and spot-size converter. (a) Transmission spectra of the two output ports of a 1 × 2 MMI and simulated optical mode profiles. (b) Transmission of the two MMI output ports as a function of MMI length. (c) Calculated transmission spectra of the eight ports of the cascaded 1 × 8 power-splitting network. (d) Top-view and side-view schematics of the SSC. (e) Three-dimensional schematic and simulated mode-field evolution in the SSC. (f) Transmission loss and transmittance of the SSC as functions of the adiabatic LN taper length.

### 2.2. Spot-Size Converter

The mode field of the TFLN rib waveguide is approximately 1 µm, which is significantly smaller than the mode field of the UHNA7 fiber. To reduce the coupling loss, a SiON-based SSC is designed, as shown in Figure 2d,e. The SSC consists of a large-mode-field SiON input waveguide, an adiabatically tapered LN waveguide, and the LN rib waveguide. The LN taper narrows in both lateral and vertical directions, enabling gradual mode transfer from the SiON waveguide to the TFLN rib waveguide. The SiON waveguide has a refractive index of approximately 1.58 at 1550 nm, between that of LN and $SiO_2$, and is designed with a 3 μm × 3 μm cross section to match the UHNA7 fiber mode. Simulation results in Figure 2f show that nearly lossless conversion is achieved when the LN taper length exceeds 300 μm. The SSC therefore serves not only as a fiber-to-chip mode transformer for the UHNA7 fiber but also as an optical interface compatible with subsequent passive butt coupling to the 1550 nm DFB laser, although further optimization of the SSC geometry is still required to minimize the DFB-to-chip coupling loss.

**2.3. High-Speed Traveling-Wave Electrodes**

To balance modulation efficiency and electrode-induced optical absorption, we first calculated the half-wave voltage-length product and the optical absorption loss as functions of the electrode gap. The results are shown in Figure 3c. A smaller electrode gap enhances the overlap between the radio-frequency electric field and the optical mode, thereby strengthening the electro-optic interaction and reducing $V_{\pi}L$. However, reducing the gap also brings the metal electrodes closer to the optical waveguide and increases optical absorption. Considering this trade-off, an electrode gap of 6.5 μm is selected, corresponding to a simulated $V_{\pi}L$ of approximately 3.0 V·cm and an absorption loss of about 0.02 dB/cm.

For high-speed Mach-Zehnder electro-optic modulators, the 3 dB modulation bandwidth is jointly determined by microwave-optical group-velocity matching, impedance matching, and microwave transmission loss[23]. In this work, a ground-signal-ground coplanar traveling-wave electrode is adopted, and these three factors are optimized in a coordinated manner. The cross section of the modulation region is shown in Figure 3a, and the simulated optical and radio-frequency electric-field distributions are shown in Figure 3b.

Microwave-optical group-velocity matching is required to maintain efficient electro-optic interaction along the full modulation length. The microwave group index can be adjusted by changing the thickness of the $SiO_2$ cladding. As shown in Figure 3d, when the $SiO_2$ cladding thickness is 2.5 μm, the microwave group index matches well with the optical group index of the TFLN waveguide, which is approximately 2.2. Impedance matching is mainly controlled by the central signal-electrode width after the electrode gap has been fixed. As shown in Figure 3e, at an electrode gap of 6.5 μm, the characteristic impedance approaches 50 Ω at high frequencies when the central signal-electrode width is optimized. A 50 Ω on-chip termination is further integrated at the electrode end to suppress microwave reflections and reduce ripples in the electro-optic frequency response.

Microwave transmission loss mainly originates from metallic ohmic loss and dielectric loss. To reduce the ohmic loss, thick Au electrodes are used. Figure 3f shows that the microwave loss decreases with increasing Au thickness, but the improvement becomes less pronounced once the Au thickness exceeds approximately 1 μm. Figures 3g-i summarize the simulated S-parameters, microwave group index, and characteristic impedance of the 7 mm traveling-wave electrode, confirming that the electrode design provides a suitable compromise among modulation efficiency, velocity matching, impedance matching, and microwave loss.

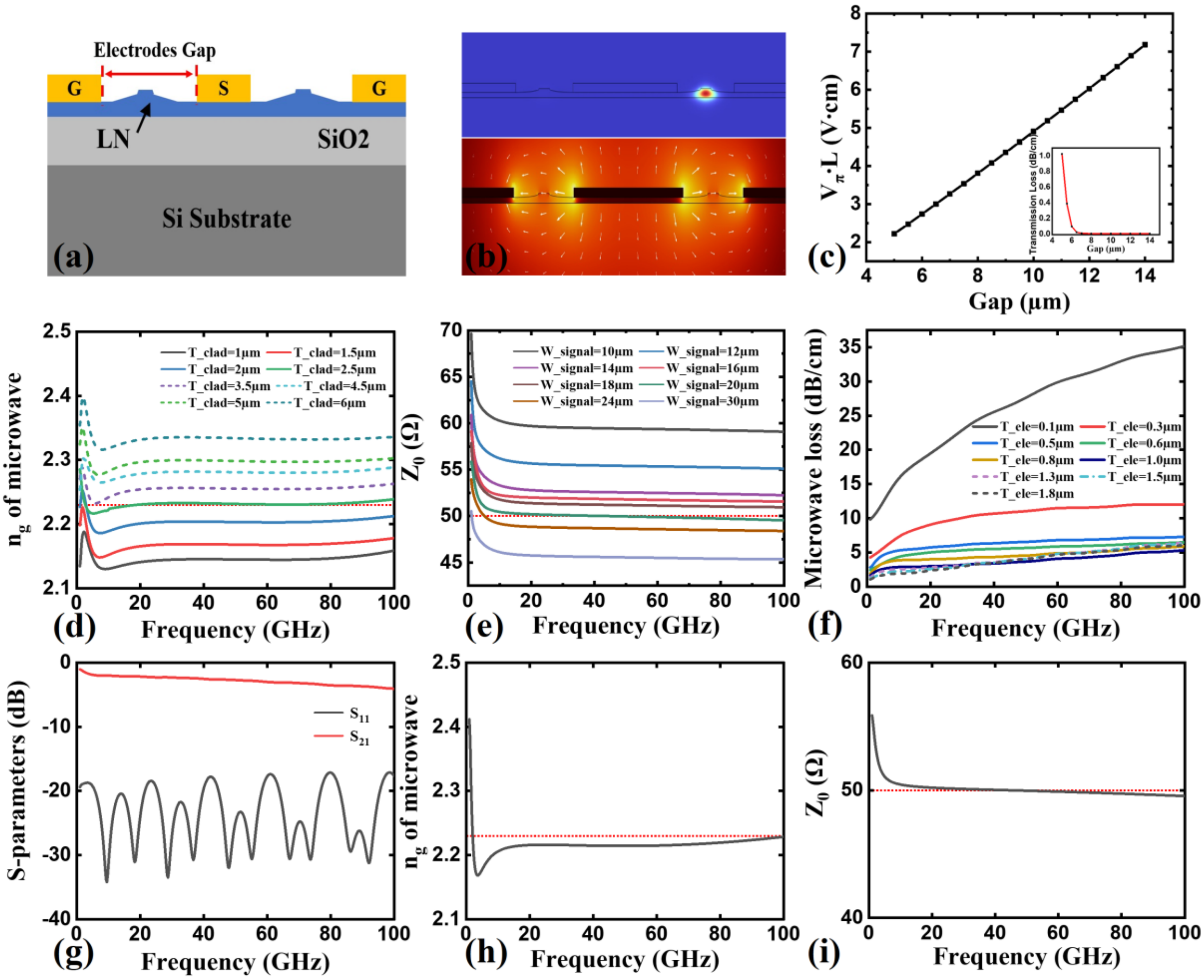

**Figure 3.** Design and simulation of the high-speed traveling-wave electrodes. (a) Cross-sectional view of the optical waveguide modulation region fabricated by the PLACE technique. (b) Simulated optical and radio-frequency electric-field distributions. (c) $V_\pi L$ and absorption loss as functions of electrode gap. (d) Microwave group refractive index versus frequency for different cladding thicknesses. (e) Characteristic impedance versus frequency for different central signal-electrode widths. (f) Microwave loss per unit length versus frequency for different electrode thicknesses. (g) Simulated S-parameters. (h) microwave group index. (i) characteristic impedance of the 7 mm traveling-wave electrode.

## 3. Fabrication

The modulator array is fabricated on a commercial x-cut lithium-niobate-on-insulator wafer consisting of a 500 nm LN thin film bonded to a buried $SiO_2$ layer on a 500 μm Si substrate. The waveguides are fabricated using femtosecond-laser-assisted chemical mechanical polishing (PLACE)[24-26]. A 200 nm Cr film is first deposited on the TFLN layer by magnetron sputtering and patterned by femtosecond laser direct writing to form a hard mask. Chemical mechanical polishing then removes the unmasked LN, and the residual Cr mask is stripped using a commercial Cr etchant.

The traveling-wave electrodes are fabricated by femtosecond laser lithography combined with wet chemical etching[27, 28]. A Au/Ti bilayer metal film is deposited on the TFLN surface. By controlling the laser power, the top Ti layer is patterned without damaging the Au layer. The Au layer is etched using aqua regia, and the Ti mask is stripped by wet etching. A $SiO_2$ cladding layer is deposited to improve microwave-optical velocity matching. Then, a 100 nm Ti layer is deposited and patterned to form the thermal tuning electrodes. Finally, a 100 nm NiCr layer is deposited and patterned to form the on-chip 50 Ω terminations. Figure 4 shows optical micrographs of the fabricated device, including the SSC, MMI, thermal tuning electrodes, traveling-wave electrodes, and termination load.

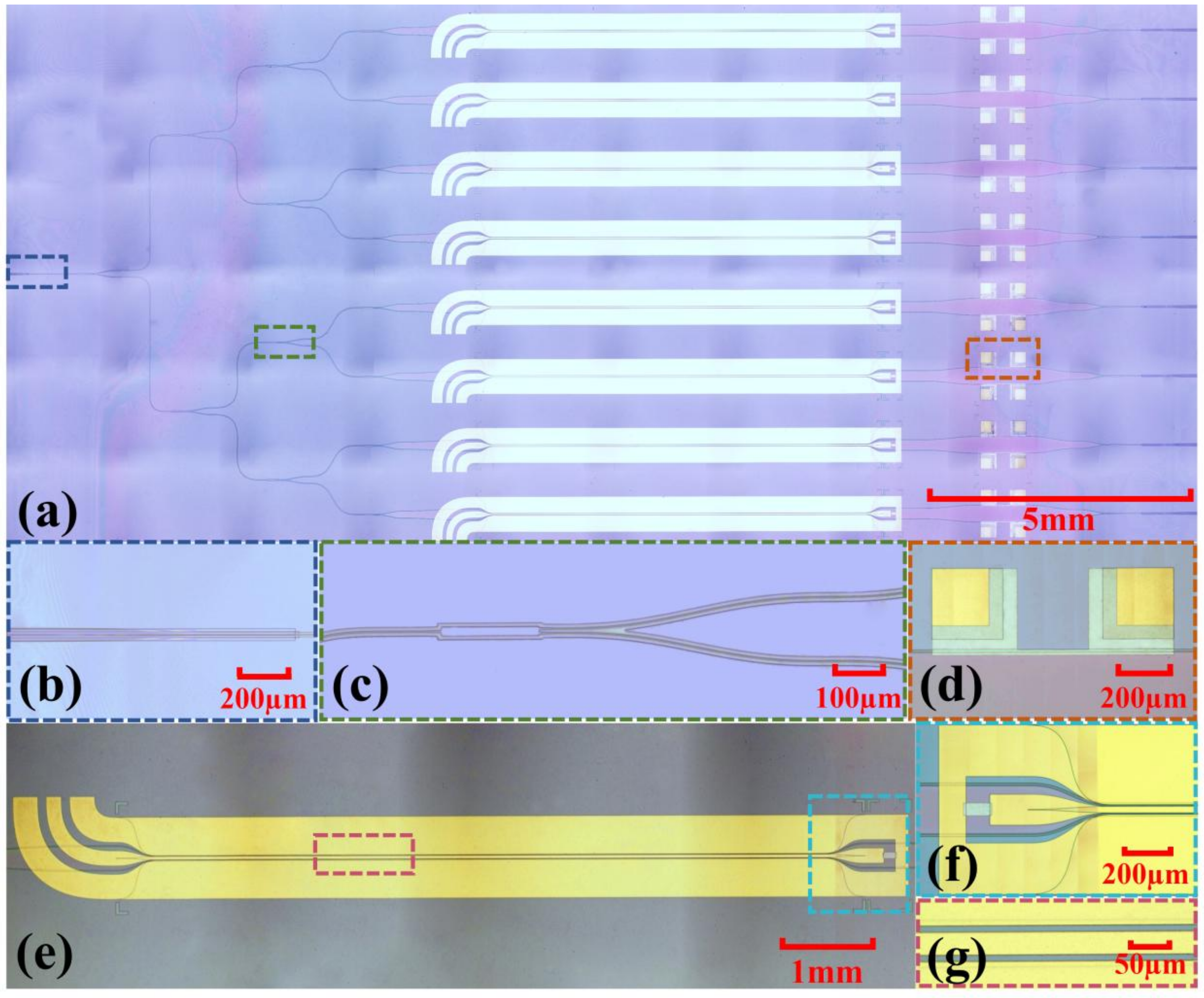

**Figure 4.** Fabricated 1 × 8 TFLN electro-optic modulator array. (a) Optical micrograph of the full chip. Zoomed-in micrographs of (b) SSC. (c) 1 × 2 MMI. (d) thermal tuning electrodes. (e) traveling-wave electrodes. (f) on-chip termination load. (g) traveling-wave modulation electrode.

## 4. Experimental Characterization

### 4.1. Power-Splitting Uniformity

The power-splitting performance of the fabricated 1 × 8 network is measured using a 1550 nm continuous-wave laser. The input polarization is adjusted to the TE mode by a polarization controller and coupled into the chip. An infrared CCD records the near-field mode distributions at the eight output ports. The optical energy at each output port is obtained by integrating the corresponding near-field intensity. As shown in Figure 5c, the splitter provides good channel-to-channel uniformity. The maximum deviation among the eight output channels is 9.7%, confirming good optical power uniformity across the eight channels.

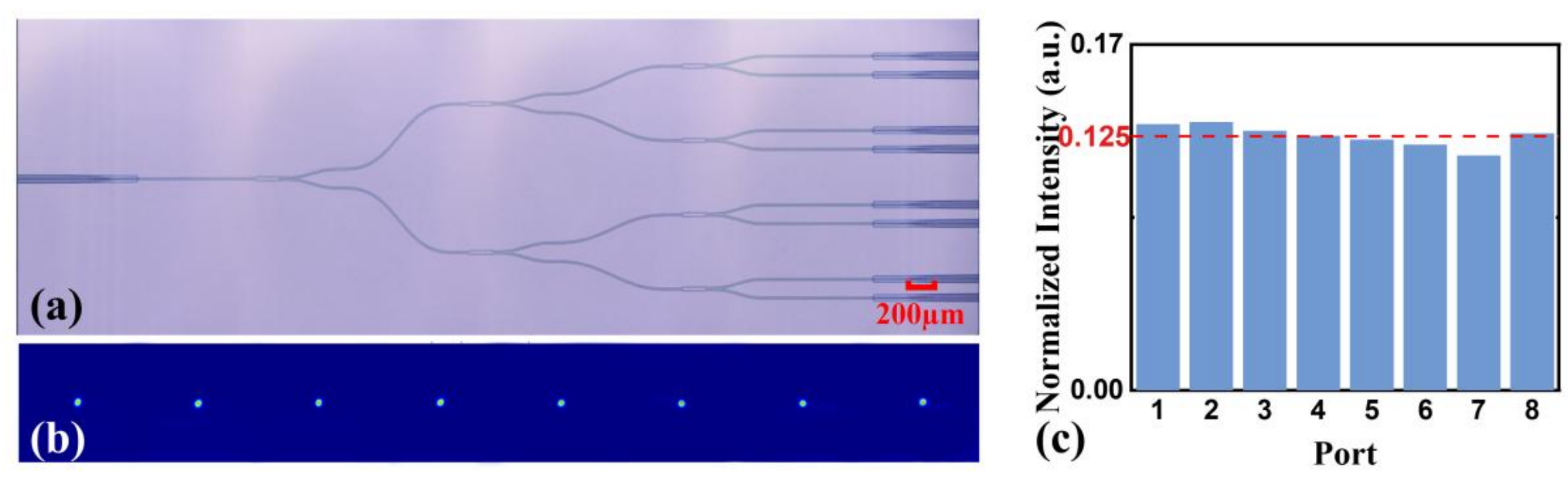

**Figure 5.** Characterization of the 1 × 8 optical power splitter. (a) Optical micrograph of the 1 × 8 MMI region. (b) Pseudo-color near-field mode distributions from the eight output ports. (c) Normalized power histogram of the eight channels.

### 4.2. Electro-Optic Frequency Response and Half-Wave Voltage

The electro-optic measurement setup is shown in Figure 6. A tunable continuous-wave laser is coupled into the chip through the input SSC using UHNA7 fiber. The output light is collected through another UHNA7 fiber and delivered to a general optical component analyzer. A vector network analyzer injects swept-frequency RF signals into each traveling-wave electrode through high-frequency coaxial cables and GSG probes. After photoelectric conversion, the RF signal is fed back to the vector network analyzer, forming a closed measurement loop for extracting the electro-optic $S_{21}$ response.

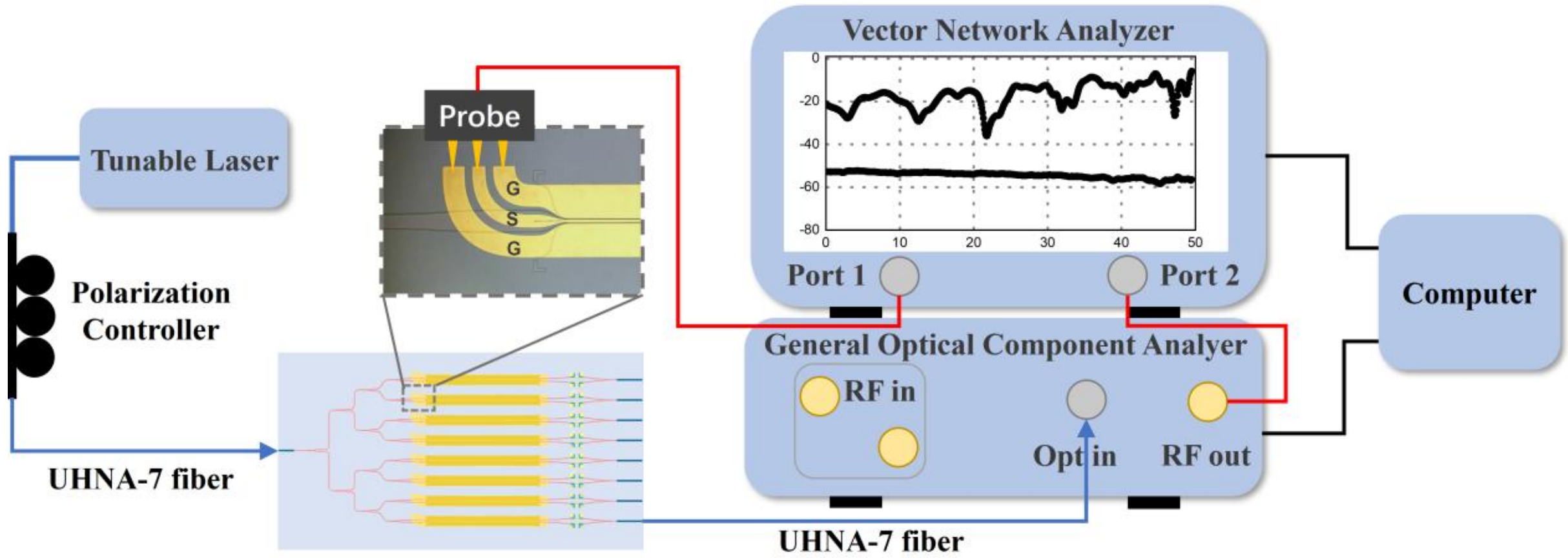

**Figure 6.** Measurement setup for the electro-optic modulator array. The optical path includes a tunable continuous-wave laser, polarization controller, UHNA7 fibers, and the TFLN modulator array. The RF path uses a vector network analyzer, high-frequency coaxial cables, GSG probes, and on-chip 50 Ω terminations.

Figure 7 presents the measured electro-optic $S_{21}$ responses of the eight channels. All channels exhibit 3 dB bandwidths exceeding 40 GHz, confirming the effectiveness of the traveling-wave electrode design. Static transfer curves are used to extract the half-wave voltages. The measured $V_\pi$ values range from 3.60 to 3.83 V. With a modulation length of 7 mm, the corresponding $V_\pi L$ are 2.52-2.68 V·cm. The inter-channel $V_\pi$ deviation is below 7%, indicating good fabrication uniformity across the modulator array. The measured extinction ratio is approximately 25 dB, indicating that the fabricated MZMs provide sufficient modulation contrast for parallel transmitter operation.

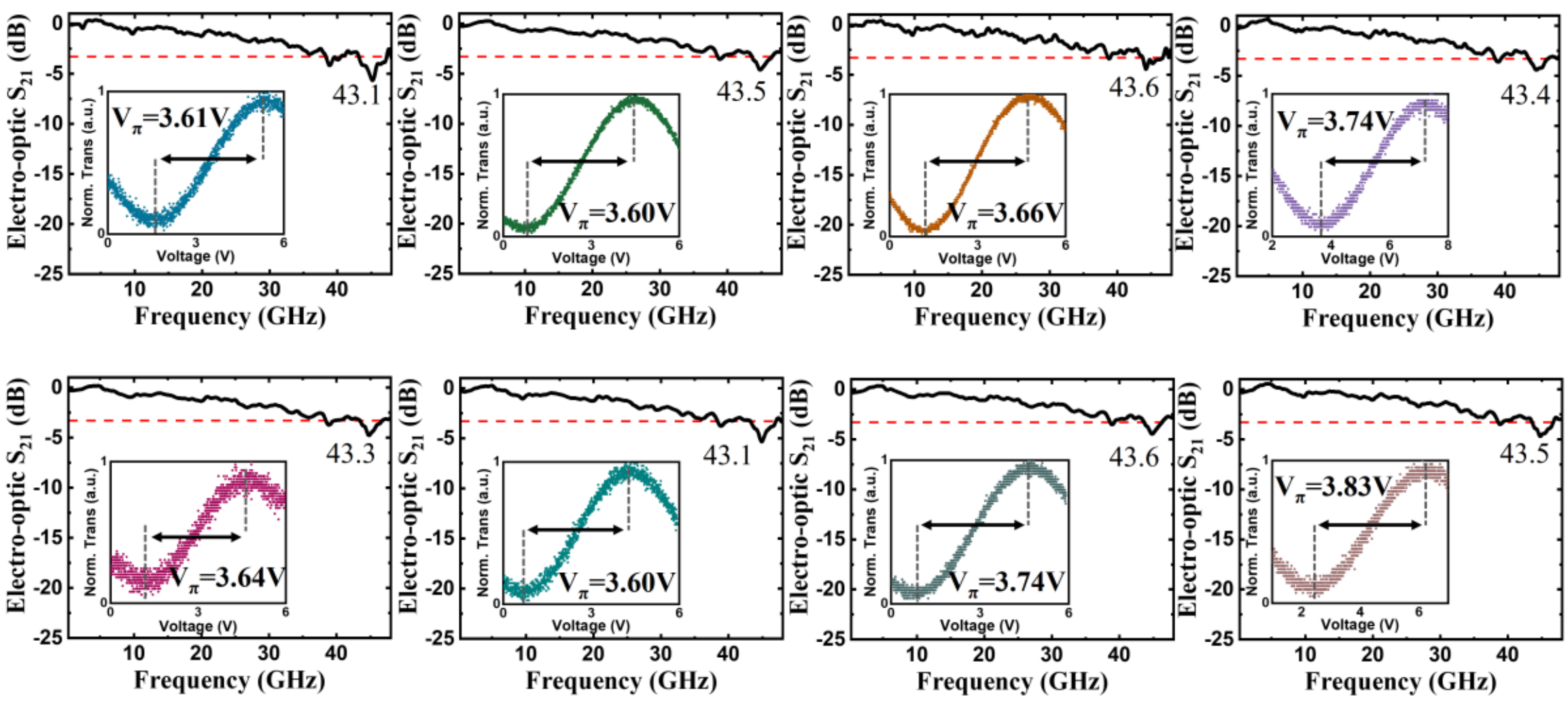

**Figure 7.** Electro-optic $S_{21}$ responses of the eight TFLN modulator channels. The red dashed line marks the -3 dB reference level. The inset shows the measured half-wave voltage $V_\pi$ of each channel.

**Table 1.** Summary of key measured and simulated performance parameters of the 1 × 8 TFLN electro-optic modulator array.

| Parameter | Value | Comment |
|---|---|---|
| Operating wavelength | C band; measured at 1550 nm | Input from external CW laser or bonded DFB laser |
| Splitter topology | Three-stage cascaded 1 × 2 MMI | 1 × 8 uniform power distribution |
| Simulated single-stage MMI loss | 0.026 dB | At 1550 nm |
| Simulated cascaded splitter loss | ≈0.084 dB | At 1550 nm |
| Electrode length | 7 mm | Traveling-wave MZM |
| Electrode gap | 6.5 μm | Compromise between $V_{\pi}L$ and absorption loss |
| Electro-optic 3 dB bandwidth | >40 GHz for all eight channels | Measured $S_{21}$ response |
| Half-wave voltage $V_{\pi}$ | 3.60-3.83 V | Static modulation curves |
| $V_{\pi}L$ | 2.52-2.68 V·cm | Calculated from measured $V_{\pi}$ and 7 mm length |
| Bare-chip insertion loss | 15.19-16.55 dB | Eight channels |
| Hybrid-integrated insertion loss | 20.19-21.55 dB | Additional DFB-to-chip coupling loss ≈5 dB |
| Inter-channel insertion-loss difference | 1.36 dB | Preserved after DFB bonding |
| Extinction ratio | ≈25 dB | Measured from static transfer curves |

### 4.3. Hybrid Integration with a DFB Laser

To evaluate the feasibility of an integrated transmitter architecture, a commercial 1550 nm DFB laser is passively butt-coupled to the TFLN chip through the input facet. Figure 8a shows the hybrid-integrated device. Figure 8b compares the insertion loss of the bare modulator array and the hybrid-integrated device. For the bare chip, the insertion loss of the eight channels ranges from 15.19 to 16.55 dB, with an inter-channel loss difference of 1.36 dB. After DFB laser bonding, the overall insertion loss increases to 20.19-21.55 dB, corresponding to an additional coupling loss of approximately 5 dB. Importantly, the inter-channel loss difference remains 1.36 dB, indicating that laser bonding does not degrade the channel uniformity of the modulator array.

The additional coupling loss is mainly attributed to mode-field mismatch between the DFB laser output and the on-chip SSC, as well as Fresnel reflection at the bonded interface. Further reduction of this loss should be possible by designing the SSC specifically for the laser near-field mode, applying facet coatings or index-matching materials, and improving passive alignment accuracy. Despite the remaining coupling loss, the present demonstration verifies a complete functional path from DFB laser input to on-chip power splitting and eight-channel parallel electro-optic modulation.

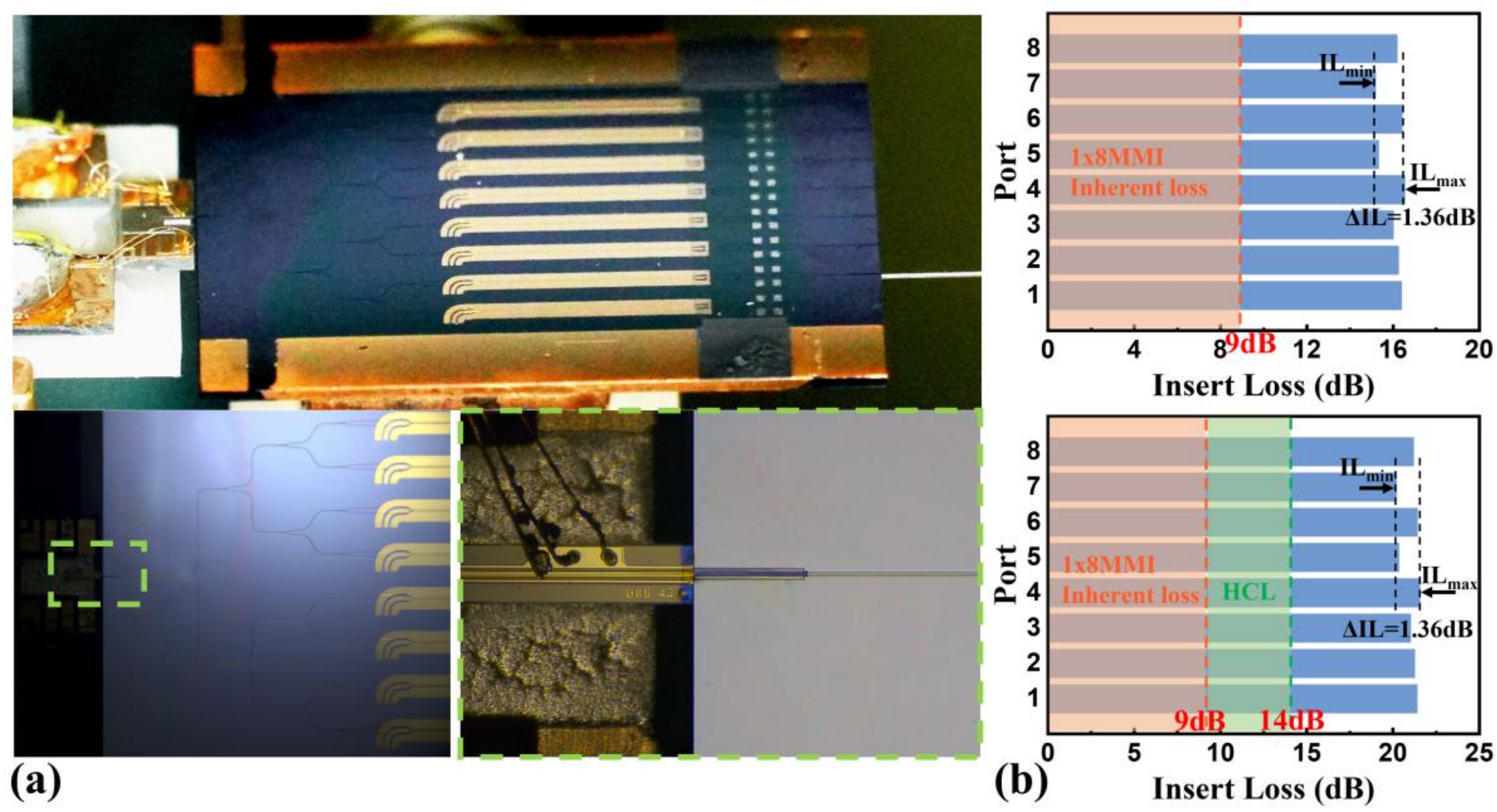

**Figure 8.** Hybrid integration and insertion-loss characterization. (a) Optical micrograph of the hybrid-integrated 8-channel high-speed electro-optic modulator-semiconductor light source. (b) Comparison of insertion loss before and after DFB laser bonding.

### 5. Conclusion and Outlook

A monolithically integrated 1 × 8 high-speed electro-optic modulator array has been designed, fabricated, and characterized on a TFLN platform. The device combines a cascaded 1 × 2 MMI power-splitting network, SSCs, eight traveling-wave MZMs, thermal tuning electrodes, and on-chip 50 Ω terminations. The splitter provides uniform optical distribution with a maximum normalized power deviation of 9.7%, while the optimized traveling-wave electrodes enable electro-optic 3 dB bandwidths above 40 GHz in all eight channels. The measured $V_\pi$ values of 3.60-3.83 V correspond to $V_\pi L$ of 2.52-2.68 V·cm, the extinction ratio reaches approximately 25 dB, and the bare-chip insertion loss remains within 15.19-16.55 dB across the array.

By passively butt-coupling a 1550 nm DFB laser to the TFLN chip, an eight-channel hybrid-integrated electro-optic light source is further demonstrated. Although the present hybrid assembly introduces approximately 5 dB of additional coupling loss, the inter-channel insertion-loss difference is preserved, confirming that the array uniformity is maintained after integration. Future work should focus on reducing the DFB-to-chip coupling loss through laser-mode-matched SSCs, facet-reflection control, and improved passive alignment. Co-packaging with driver electronics and thermal control will also be needed for practical deployment. Overall, this work establishes a scalable TFLN modulator-array prototype and provides a technically feasible route toward compact optical transmitters for high-speed parallel interconnects.

**Acknowledgements**

This research was funded by National Key R&D Program of China（2025YFF0524600). National Natural Science Foundation of China(12192251, 12334014, 62335019, 12134001, 12304418, 12474378). Quantum Scienceand Technology-National Science and Technology Major Project (2021ZD0301403). Shanghai Municipal Science and Technology Major Project (GrantNo.2019SHZDZX01)

Received: ((will be filled in by the editorial staff))

Revised: ((will be filled in by the editorial staff))

Published online: ((will be filled in by the editorial staff))